\documentclass[12pt]{article}
\usepackage[dvips]{color}
\usepackage{amsmath}
\usepackage{graphicx}
\usepackage{subfigure}
\usepackage{epsfig}
\usepackage{amsmath}
\usepackage{cite}
\usepackage{color}
\usepackage{subfigure}

\begin{document}
\begin{center}
\Large{\bf Anisotropic constant-roll inflationary scenario with complex quintessence field and swampland conjectures }\\
\small \vspace{1cm} {\bf Jafar Sadeghi$^{\star}$\footnote {Email:~~~pouriya@ipm.ir}}, \quad
{\bf Saeed Noori Gashti$^{\star}$\footnote {Email:~~~saeed.noorigashti@stu.umz.ac.ir}}, \quad
{\bf Mehdi Shokri$^{\dagger,\ddagger}$\footnote {Email:~~~mehdishokriphysics@gmail.com}}
\\
\vspace{0.5cm}$^{\star}${Department of Physics, Faculty of Basic
Sciences,\\
University of Mazandaran
P. O. Box 47416-95447, Babolsar, Iran}\\
$^{\dagger}${Department of Physics, University of Tehran, North Karegar Ave., Tehran 14395-547, Iran}\\
$^{\ddagger}${School of Physics, Damghan University, P. O. Box 3671641167, Damghan, Iran}
\small \vspace{1cm}
\small \vspace{1cm}
\end{center}
\begin{abstract}
In this paper, with presence of swampland conjecture  we use the complex form of a scalar field and investigate the anisotropic constant-roll of an inflationary scenario.  Recently, the  complex quintessence field has been used to describe the accelerating expansion of the universe which has had interesting results connecting with various conditions. Therefore, the Lagrangian density of the quintessence field leads to obtain the  equations of the complex scalar field. Also,  we use the anisotropic constant-roll conditions and the field equation and  calculate the exact solutions for some cosmology parameters such as Hubble parameter and potential.
Using the exact solution of potential and refined swampland conditions, we plot some figures. In that case, the figures lead us to have challenge between three concepts as  complex quintessence, anisotropic constant-roll conditions and swampland conjectures. Then we discuss their compatibility and incompatibility with corresponding results. Finally, we will analyze the complex quintessence in examining the inflationary scenario.
\end{abstract}
\newpage
\tableofcontents

\section{Introduction}

Recently, reports from various projects such as High-Z Supernova Search and Supernova Cosmology Project about their observations of SNe Ia have shown that the universe has an accelerating cosmic expansion \cite{1,2}.
In addition, the CMB measurements from MAXIMA-1 and BOOMERANG represent the fact that there is around $l=200$ a sharp zenith, which indicates that the universe is flat \cite{3,4}.
Of course, multiple projects have made various observations so far, each for a specific mission, to use such cosmic data to identify the large-scale structures and nature of the universe and how this universe formed.
The above observations suggest that the universe may have a critical density that could include different structures, representing one-third of ordinary matter and two-thirds of dark energy with negative pressure\cite{5,6}.\\
 So far, various concepts have been used to explain the universe's acceleration. One of the most important candidates is dark energy with unknown nature. Due to this unknown nature also introduced different types of models including positive cosmological constant, quintessence, Chaplygin gas models, interacting dark energy models, ghost condensate, quintom, tachyon models and braneworld models \cite{7,8}
 One of the models we mentioned is the quintessence, which includes different parts (real and complex).
Huterer, Turner, and other researchers have conducted studies on the real part of the quintessence field\cite{9,10,11,12}.
Also, the complex part of the quintessence field has recently been considered and discussed in dark energy structures. Sadeghi Noori Gashti and Azizi studied Tsallis and Kaniadakis holographic dark energy with Complex Quintessence theory in Brans- Dicke cosmology and Tsallis holographic dark energy under the Complex form of Quintessence model\cite{13,14}, also for the further reading you see \cite{15}.\\
Another candidate for examining the universe's accelerating expansion is the inflationary scenario,
 which has been studied from different models and conditions so far.
One of the most important conditions in which much work has been done is studying the inflation model concerning the constant-roll condition. Constant- roll condition is as $\ddot{\phi}=\beta H\dot{\phi}$.
 The exact solutions calculated in constant-roll conditions can be closer to the latest observable data constant-roll\cite{16,17,18,19,20}
This condition has recently been investigated in various theories such as teleparallel $f (t)$ gravity, $f (R)$, $f (T),$ $f(Q)$, etc., and the results have been compared with the latest observable data.
Also, the anisotropic constant-roll condition has recently been studied by researchers in various inflation models, which has led to exciting results\cite{21}.\\
Recently, researchers have used a combination of this theory with other conditions and conjectures, including the condition derived from string theory, called the swampland program. The swampland program has have been studied in various cosmological structures, including inflation, dark energy, and the physics of black holes.  Despite the problems and novelty of this idea, it has also yielded interesting results.
For further study of this concept, as well as the most important conjectures that have been studied in this theory, such as dS conjecture, weak gravity conjecture, etc., see \cite{22,23,24,25,26,27,28,29,30,31,32,33,34,35,36}.
In this paper, we want to study the constant-roll condition of the inflationary scenario in the presence of refined swampland conjectures considering the complex form of the scalar field as complex quintessence. We compare the obtained  results with the latest observable data and other studies. This concept has not been studied yet and can lead to exciting results.
 According to the concept, as mentioned earlier, the organization of this article is as follows.\\
 In section 2 we describe the concept of complex quintessence field and basic equations by introducing the resulting field equations.
In section 3, we apply the anisotropic constant-roll condition concerning the complex form of the scalar field or the complex quintessence and obtain accurate solutions, including the Hubble parameter and the potential.
In section 4, according to the exact solutions that we calculated for the potential of the mentioned model, we apply the refined swampland condition and describe the results by plotting some figures and challenge the compatibility or incompatibility of the mentioned concepts. Finally, we describe the results in section 5.
\section{Complex Quintessence theory}
In this section, we first state the basic equations for studying anisotropic constant-roll conditions. So we will first assume that a flat spatial universe is dominated by non-relativistic matter and a complex scalar field, which can be expressed with the following flat metric\cite{39},
\begin{equation}\label{1}
\begin{split}
ds^{2}=dt^{2}-a^{2}(t)(dr^{2}+r^{2}d\varphi_{1}^{2}+r^{2}\sin^{2}\varphi_{1}d\varphi_{2}^{2}),
\end{split}
\end{equation}
In the second step, we  introduce the corresponding action, which is given by,
\begin{equation}\label{2}
\begin{split}
S=\int d^{4}x\sqrt{g}\big(-\frac{1}{16\pi G}\mathcal{R}-\rho_{M}+\mathcal{L}_{\Phi}\big),
\end{split}
\end{equation}
where  $g$, $G$, $\mathcal{R}$ and $\rho_{M}$ are the absolute value of the metric tensor $g_{\mu\nu}$, Newton’s constant, and Ricci scalar and matter density, respectively.
According to the last term in equation (2), the Lagrangian density for this complex scalar field is expressed by,
\begin{equation}\label{3}
\begin{split}
\mathcal{L}_{\Phi}=\frac{1}{2}g^{\mu\nu}\partial_{\mu}\Phi^{\ast}\partial_{\nu}\Phi-V(|\Phi|),
\end{split}
\end{equation}
where $\mu, \nu=0,1,2,3$. In the Lagrange equation, we will assume that the potential term depends only on the exact values of the complex form of the scalar field. In the following, instead of the two field parameters $\Phi$ and $\Phi^{\ast}$ we will use alternative field variables such as $\phi$ and $\theta$ called the amplitude and the phase, respectively. So, we have do separation $\Phi$ and $\Phi^{\ast}$  in terms of $\phi$ and $\theta$,  we will have following,
\begin{equation}\label{4}
\begin{split}
\Phi(x)=\phi(x)\exp(i\theta x),
\end{split}
\end{equation}
Or the more accurate form is as follows.
\begin{equation}\label{5}
\begin{split}
\Phi(t)=\phi(t)\exp(i\theta t),
\end{split}
\end{equation}
According to the new  definition of variables, the Lagrangian equation (3) is rewritten as follows,
\begin{equation}\label{6}
\begin{split}
\mathcal{L}_{\Phi}=\frac{1}{2}g^{\mu\nu}\partial_{\mu}\phi\partial_{\nu}\phi+\frac{1}{2}
\phi^{2}g^{\mu\nu}\partial_{\mu}\theta\partial_{\nu}\theta-V(\phi),
\end{split}
\end{equation}
Now with the help of equations (2) and (6), we can calculate Einstein and field equations of the complex scalar field. Therefore, using the metric equation (1), the above mentioned equations are obtained in the following form,
\begin{equation}\label{7}
\begin{split}
H^{2}\equiv(\frac{\dot{a}}{a})^{2}=\frac{8\pi G}{3}\rho=\frac{8\pi G}{3}\bigg(\rho_{M}+\big(\frac{1}{2}\dot{\phi}^{2}+\frac{1}{2}\phi^{2}\dot{\theta}^{2}\big)+V(\phi)\bigg)
\end{split}
\end{equation}
\begin{equation}\label{8}
\begin{split}
(\frac{\ddot{a}}{a})=-\frac{4\pi G}{3}(\rho+3p)=-\frac{8\pi G}{3}\bigg(\frac{1}{2}\rho_{M}+\big(\dot{\phi}^{2}+\phi^{2}\dot{\theta}^{2}\big)-V(\phi)\bigg),
\end{split}
\end{equation}
\begin{equation}\label{9}
\begin{split}
\ddot{\phi}+3H\dot{\phi}-\dot{\theta}^{2}\phi+V'(\phi)=0,
\end{split}
\end{equation}
\begin{equation}\label{10}
\begin{split}
\ddot{\theta}+\big(2\frac{\dot{\phi}}{\phi}+3H\big)\dot{\theta}=0,
\end{split}
\end{equation}
The above equations are the fundamental equations that govern the universe's evolution
and  $H$, $'$ $dot$ are Hubble parameters, derivatives concerning $\phi$, and derivatives with respect to $t$, respectively.
Also,  two parameters $\rho$ and $p$ are called energy density and pressure.
It would be worth mentioning here that the contribution is non-relativistic matter is energy density and pressure $\rho_{M}=p_{M}=0$. In that case, energy density and pressure is expressed in the following form.
\begin{equation}\label{11}
\begin{split}
\rho_{\Phi}=\frac{1}{2}\big(\dot{\phi}^{2}+\phi^{2}\dot{\theta}^{2}\big)+V(\phi)\\
p_{\Phi}=\frac{1}{2}\big(\dot{\phi}^{2}+\phi^{2}\dot{\theta}^{2}\big)-V(\phi),
\end{split}
\end{equation}
According to the solution of equation (10), a relation for "angular velocity" is calculated in the following form.
\begin{equation}\label{12}
\begin{split}
\dot{\theta}=\frac{\omega}{a^{3}\phi^{2}},
\end{split}
\end{equation}
where $\omega$ is an integration constant calculated according to the initial conditions defined for $\dot{\theta}$. According to the equation derived from (12), equations (7-10) can be rewritten as follows.
\begin{equation}\label{13}
\begin{split}
H^{2}\equiv(\frac{\dot{a}}{a})^{2}=\frac{8\pi G}{3}\rho=\frac{8\pi G}{3}\bigg(\rho_{M}+\frac{1}{2}\dot{\phi}^{2}+\frac{1}{2}\frac{\omega^{2}}{a^{6}\phi^{2}}+V(\phi)\bigg)
\end{split}
\end{equation}
\begin{equation}\label{14}
\begin{split}
(\frac{\ddot{a}}{a})=-\frac{4\pi G}{3}(\rho+3p)=-\frac{8\pi G}{3}\bigg(\frac{1}{2}\rho_{M}+\dot{\phi}^{2}+\frac{\omega^{2}}{a^{6}\phi^{2}}-V(\phi)\bigg),
\end{split}
\end{equation}
\begin{equation}\label{15}
\begin{split}
\ddot{\phi}+3H\dot{\phi}-\frac{\omega^{2}}{a^{6}\phi^{3}}+V'(\phi)=0,
\end{split}
\end{equation}

...........As mentioned earlier, the term  $\frac{\omega^{2}}{2a^{6}\phi^{2}}$ is derived from the angular motion of the complex form of the scalar field, which can also be referred to as an effective potential. In fact, in a more severe form, perhaps the centrifugal potential causes a distance $\phi$ from zero by creating a centrifugal force.
Also, since the gravitational effect of dark energy with negative pressure prevents the ordinary material from forming configurations, the contributions related to quintessence to pressure and energy density have been negligible for a short time ago. Also, to maintain the balance between different theories such as Big Bang Cosmology + Inflation + Cold Dark Matter and observations as measurements of CMB, quintessence field contributions at various approaches such as recombination and primordial nucleosynthesis should be insignificant.
These constraints on quintessence help distinguish between two types of kinetic energy resulting from the complex form of the scalar field or (complex quintessence). As shown in the fundamental equations (13) and (14), the contribution of the angular motion of the scalar complex field to the energy density and the pressure is proportional to the expression $a^{-6}\phi^{-2}$.
Of course, factor $a^{-6}$ may cause a rapid decrease in these shares, although it is also subject to conditions; for example, $\phi$ does not decrease with $(a^{-3/2})$.
For more study, you can see \cite{39}.
Also, according to equations (13-15), we can implement the anisotropic constant-roll condition. However, according to the data $r(z)$ from SNe Ia and some parameters such as $\Omega_{M}$, $H_{0}$, and $\omega$, the results we can discuss in detail.
Of course, the most important parameters, $z$,  $r(z)$, $H_{0}$, and $\Omega_{M}$ are redshift, Robertson-Walker coordinate distance, Hubble constant, and matter-energy density fraction. we can also benefit two equations as $1+z =1/a$ where $a$ named the scale factor and $\rho_{M}=\Omega_{M}\rho_{c}=\frac{3\Omega_{M}H_{0}^{2}}{8\pi G}(1+z)^{3}$

\section{Anisotropic constant-roll inflation}

This section will discuss the anisotropic constant-roll condition concerning the complex form of the scalar field, such as complex quintessence. So we can obtain accurate solutions for parameters such as Hubble and potential, i.e., parameters that play an important role in explaining the inflation scenario.
The basic condition that we will assume here is that the first index of slow-roll condition in the inflation period is very small. Therefore, according to all concepts mentioned earlier and according to equations (13) and (14), we will have,

\begin{equation}\label{16}
\begin{split}
3\frac{\ddot{a}}{a}-3H^{2}=-\frac{3}{2}\rho_{M}-\frac{3}{2}\dot{\phi}^{2}-\frac{3}{2}\frac{\omega^{2}}{a^{6}\phi^{2}},
\end{split}
\end{equation}

According to the definition of the Hubble parameter $H=\frac{\dot{a}}{a}$ and equation (16), one can calculate,

\begin{equation}\label{17}
\begin{split}
\dot{H}=-\frac{3\Omega_{M}H_{0}^{2}}{2a^{3}}-\frac{1}{2}\dot{\phi}^{2}-\frac{1}{2}\frac{\omega^{2}}{a^{6}\phi^{2}},
\end{split}
\end{equation}

We use the auxiliary equation viz $\dot{H}=\frac{dH}{d\phi}\dot{\phi}$ and place it in the above equation. so we will have,

\begin{equation}\label{18}
\begin{split}
\frac{1}{2}\dot{\phi}^{2}+\frac{dH}{d\phi}\dot{\phi}+\frac{1}{2}\frac{\omega^{2}}{a^{6}\phi^{2}}+\frac{3\Omega_{M}H_{0}^{2}}{2a^{3}},
\end{split}
\end{equation}

We solve the above equation in terms of $\dot{\phi}$, So

\begin{equation}\label{19}
\begin{split}
\dot{\phi}=-\frac{dH}{d\phi}\pm\sqrt{\frac{d^{2}H}{d\phi^{2}}-(\frac{\omega^{2}}{a^{6}\phi^{2}}+\frac{3\Omega_{M}H_{0}^{2}}{a^{3}})},
\end{split}
\end{equation}

We take the derivative from the above equation concerning $t$. so we will have,

\begin{equation}\label{20}
\begin{split}
\ddot{\phi}=-\frac{d^{2}H}{d\phi^{2}}\dot{\phi}\pm\frac{\frac{d^{2}H}{d\phi^{2}}\frac{dH}{d\phi}\dot{\phi}-(\frac{-2\omega^{2}\dot{\phi}}{a^{6}\phi^{3}})}{\sqrt{\frac{d^{2}H}{d\phi^{2}}-(\frac{\omega^{2}}{a^{6}\phi^{2}}+\frac{3\Omega_{M}H_{0}^{2}}{a^{3}})}},
\end{split}
\end{equation}

We combine equation (20) with $\ddot{\phi}=-(3+\alpha)H\dot{\phi}$. So we will have.

\begin{equation}\label{21}
\begin{split}
-(3+\alpha)H=-\frac{d^{2}H}{d\phi^{2}}\pm\frac{\frac{d^{2}H}{d\phi^{2}}\frac{dH}{d\phi}-(\frac{-2\omega^{2}}{a^{6}\phi^{3}})}{\sqrt{\frac{d^{2}H}{d\phi^{2}}-(\frac{\omega^{2}}{a^{6}\phi^{2}}+\frac{3\Omega_{M}H_{0}^{2}}{a^{3}})}},
\end{split}
\end{equation}

Assuming $\omega=\Omega_{M}=0$ the above equation becomes as,

\begin{equation}\label{22}
\begin{split}
\frac{d^{2}H}{d\phi^{2}}-\frac{3+\alpha}{2}H=0,
\end{split}
\end{equation}

The solution of the above-reduced equation is calculated as follows.

\begin{equation}\label{23}
\begin{split}
H=c_{1}\exp(\sqrt{\frac{3+\alpha}{2}}\phi)+c_{2}\exp(-\sqrt{\frac{3+\alpha}{2}}\phi),
\end{split}
\end{equation}

The above equation is a specific solution of the Hubble parameter or the general solution of the isotropic constant-roll condition. We assumed the effects of specific parameters in this equation to be zero.
According to equation (23), a particular ansatz can be considered for the general solution in equation (21) in the following form to specify the effects of the deleted parameters.

\begin{equation}\label{24}
\begin{split}
H=c_{1}\exp(\lambda(\omega, \Omega_{M})\sqrt{\frac{3+\alpha}{2}}\phi)+c_{2}\exp(-\lambda(\omega, \Omega_{M})\sqrt{\frac{3+\alpha}{2}}\phi),
\end{split}
\end{equation}

Here, by placing this ansatz in the equation (21) and with a series of direct calculations, the parameter $\lambda$ can be specified.
We compute the parameter $\lambda$ and place it in the Hubble equation (24). We then examine specific solutions using boundary conditions.
We will also calculate the exact solutions for essential parameters, such as the potential for investigating swampland conjectures.

\subsection{Condition I}

Here we will apply the boundary condition as $(c_{1}=c_{2}=M/2)$ to the Hubble parameter in equation (24).
Since different values are calculated for the Hubble parameter, we kept only positive solutions.
The negative solutions showed a similar concept in the study of the swampland program.
Hence, we only pursue positive solutions.
Negative solutions follow a similar process so that we will have.

\begin{equation}\label{25}
\begin{split}
H=M\cosh\bigg[\sqrt{\frac{3+\alpha}{2}}\phi\bigg(\pm\frac{2\sqrt{M(3+\alpha)-\frac{2\omega^{2}}{a^{6}\phi^{3}\sqrt{1/2M(3+\alpha)-12H_{0}^{2}\Omega_{M}/a^{3}}}}}{\sqrt{\frac{M\big\{-12H_{0}^{2}(3+\alpha)\Omega_{M}+a^{3}M\big[2+(3+\alpha)^{2}+\phi\sqrt{2M(3+\alpha)-12H_{0}^{2}\Omega_{M}/a^{3}}\big]\big\}}{a^{3}M(3+\alpha)-6H_{0}^{2}\Omega_{M}/a^{3}}}}\bigg)\bigg],
\end{split}
\end{equation}

Given the relation obtained for the Hubble parameter in equation (25), we will have

\begin{equation}\label{26}
\begin{split}
&\mathcal{X}_{1}=\sqrt{M(3+\alpha)-\frac{2\omega^{2}}{a^{6}\phi^{3}\sqrt{\frac{1}{2}M(3+\alpha)-3H_{0}^{2}\Omega_{M}/a^{3}}}}\\
&\mathcal{X}_{2}=\sqrt{\frac{M\big(-12H_{0}^{2}(3+\alpha)\Omega_{M}+a^{3}M\big[2(3+\alpha)^{2}+\phi\sqrt{2M(3+\alpha)-12H_{0}^{2}\Omega_{M}/a^{3}}\big]\big)}{a^{3}M(3+\alpha)-6H_{0}^{2}\Omega_{M}/a^{3}}}\\
&\mathcal{X}_{3}=a^{2}\phi\sqrt{2M(3+\alpha)-12H_{0}^{2}\Omega_{M}/a^{3}}\\
&V(\phi)=-2M^{2}\bigg(-\frac{a^{3}M^{2}\sqrt{3+\alpha}\mathcal{X}_{1}\times\mathcal{X}_{3}}{\sqrt{2}(a^{3}M(3+\alpha)-6H_{0}^{2}\Omega_{M}/a^{3})(\mathcal{X}_{2})^{3}}+\frac{3\sqrt{2}\sqrt{3+\alpha}\omega^{2}}{a^{3}\phi^{2}\mathcal{X}_{1}\times\mathcal{X}_{2}\times\mathcal{X}_{3}}\\
&+\frac{\sqrt{2}\sqrt{3+\alpha}\mathcal{X}_{1}}{\mathcal{X}_{2}}\bigg)^{2}\times\sinh^{2}\big[\frac{\sqrt{2}\sqrt{3+\alpha}\phi\mathcal{X}_{1}}{\mathcal{X}_{2}}\big]+\cosh^{2}\big[\frac{\sqrt{2}\sqrt{3+\alpha}\phi\mathcal{X}_{1}}{\mathcal{X}_{2}}\big],
\end{split}
\end{equation}

Also, according to equations (19) and (25), we can calculate $\dot{\phi}$. So we will have,

\begin{equation}\label{27}
\begin{split}
&\mathcal{X}_{1}=\sqrt{M(3+\alpha)-\frac{2\omega^{2}}{a^{6}\phi^{3}\sqrt{\frac{1}{2}M(3+\alpha)-3H_{0}^{2}\Omega_{M}/a^{3}}}}\\
&\mathcal{X}_{2}=\sqrt{\frac{M\big(-12H_{0}^{2}(3+\alpha)\Omega_{M}+a^{3}M\big[2(3+\alpha)^{2}+\phi\sqrt{2M(3+\alpha)-12H_{0}^{2}\Omega_{M}/a^{3}}\big]\big)}{a^{3}M(3+\alpha)-6H_{0}^{2}\Omega_{M}/a^{3}}}\\
&\mathcal{X}_{3}=a^{2}\phi\sqrt{2M(3+\alpha)-12H_{0}^{2}\Omega_{M}/a^{3}}\\
&\dot{\phi}=-M\bigg(-\frac{a^{3}M^{2}\sqrt{3+\alpha}\mathcal{X}_{3}\mathcal{X}_{1}}{\sqrt{2}(a^{3}M(3+\alpha)-6H_{0}^{2}\Omega_{M}/a^{2})(\mathcal{X}_{2})^{3}}+\frac{3\sqrt{2}\sqrt{3+\alpha}\omega^{2}}{a^{3}\phi^{2}\mathcal{X}_{1}\times\mathcal{X}_{2}\times\mathcal{X}_{3}}\\
&+\frac{\sqrt{2}\sqrt{3+\alpha}\mathcal{X}_{1}}{\mathcal{X}_{2}}\bigg)\sinh\big[\frac{\sqrt{2}\sqrt{3+\alpha}\phi\mathcal{X}_{1}}{\mathcal{X}_{2}}\big]-\bigg(M\bigg[-\frac{a^{3}M^{2}\sqrt{3+\alpha}\mathcal{X}_{3}\times\mathcal{X}_{1}}{\sqrt{2}(a^{3}M(3+\alpha)-6H_{0}^{2}\Omega_{M}/a^{3})/(\mathcal{X}_{2})^{3}}\\
&+\frac{3\sqrt{2}\sqrt{3+\alpha}\omega^{2}}{a^{3}\phi^{2}\mathcal{X}_{1}\times\mathcal{X}_{2}\times\mathcal{X}_{3}}+\frac{\sqrt{2}\sqrt{3+\alpha}\mathcal{X}_{1}}{\mathcal{X}_{2}}\bigg]^{2}-\frac{\omega^{2}}{a^{6}\phi^{2}}-3H_{0}^{2}\Omega_{M}/a^{3}\\
&\times\cosh\big[\frac{\sqrt{2}\sqrt{3+\alpha}\phi\mathcal{X}_{1}}{\mathcal{X}_{2}}\big]+M\bigg\{\frac{3a^{3}M^{4}\sqrt{3+\alpha}\phi(2M(3+\alpha)-12H_{0}^{2}\Omega_{M}/a^{3})\mathcal{X}_{1}}{2\sqrt{2}(a^{3}M(3+\alpha)-6H_{0}^{2}\Omega_{M}/a^{3}(\mathcal{X}_{2})^{5}}\\
&-\frac{3\sqrt{2}M^{2}\sqrt{3+\alpha}\omega^{2}\mathcal{X}_{3}}{\phi^{2}(a^{3}M(3+\alpha)-6H_{0}^{2}\Omega_{M}\mathcal{X}_{3}\times\mathcal{X}_{1}\times(\mathcal{X}_{2})^{3}}-\frac{\sqrt{2}a^{3}M^{2}\sqrt{3+\alpha}\mathcal{X}_{3}\times\mathcal{X}_{1}}{(a^{3}M(3+\alpha)-6H_{0}^{2}\Omega_{M}/a^{3})(\mathcal{X}_{2})^{3}}\\
&-\frac{6\sqrt{2}\sqrt{3+\alpha}\omega^{2}}{a^{3}\phi^{3}\mathcal{X}_{1}\times\mathcal{X}_{2}\times\mathcal{X}_{3}}-\frac{9\sqrt{2}\sqrt{3+\alpha}\omega^{4}}{a^{6}\phi^{6}(1/2 M(3+\alpha)-3H_{0}^{2}\Omega_{M}/a^{3})(\mathcal{X}_{1})^{3}\mathcal{X}_{2}}\bigg\}\\
&\times\sinh\big[\frac{\sqrt{2}\sqrt{3+\alpha}\phi\mathcal{X}_{1}}{\mathcal{X}_{2}}\big]\bigg),
\end{split}
\end{equation}

Different solutions can be calculated for the parameter $\phi$ by solving the above equation.
Also, we can obtain with straightforward calculation the scale factor according to equations (25) and $H=\frac{\dot{a}}{a}$. We can examine another boundary condition and compare the results with each other.

\subsection{Condition II}

In this part, we consider another boundary condition expressed as $c_{1}=M/2$ and $c_{2}=-M/2$. Also, we focus only on positive solutions. Therefore, according to the computational values for the parameter $\lambda$, we will calculate the quantities as Hubble, potential, and $\dot{\phi}$ separately. Hence we will have

\begin{equation}\label{28}
\begin{split}
H=M\sinh\bigg[\sqrt{\frac{3+\alpha}{2}}\phi\bigg(\pm\frac{\sqrt{\frac{M^{2}(3+\alpha)\bigg\{-12H_{0}^{2}(3+\alpha)^{2}\Omega_{M}+\frac{2a^{3}M\sqrt{3+\alpha}\omega^{2}}{(3+\alpha)^{3}\phi+3\sqrt{\frac{\sqrt{2}a^{3}M(3+\alpha)^{3/2}\phi-3H_{0}^{2}\Omega_{M}}{a^{3}}}}\bigg\}}{\sqrt{2}a^{3}M(3+\alpha)^{3/2}\phi-12H_{0}^{2}\Omega_{M}/a^{3}}}}{M(3+\alpha)^{3/2}\phi}\bigg)\bigg],
\end{split}
\end{equation}

Here, as in the previous part, quantities such as potential and $\dot{\phi}$ will be calculated as follows.

\begin{equation}\label{29}
\begin{split}
&\mathcal{Y}_{1}=\bigg(\frac{M^{2}(3+\alpha)\bigg[-12H_{0}^{2}(3+\alpha)^{2}\Omega_{M}+\frac{2a^{3}M\sqrt{3+\alpha}\omega^{2}}{(3+\alpha)^{3}\phi+3\sqrt{\frac{\sqrt{2}a^{3}M(3+\alpha)^{2}-3H_{0}^{2}\Omega_{M}}{a^{3}}}}\bigg]}{\sqrt{2}a^{3}M(3+\alpha)^{3/2}\phi-12H_{0}^{2}\Omega_{M}/a^{3}}\bigg)^{\frac{1}{2}}\\
&\mathcal{Y}_{2}=\big((\sqrt{2}a^{3}M(3+\alpha)^{2}-3H_{0}^{2}\Omega_{M})/a^{3}\big)^{1/2}\\
&\mathcal{Y}_{3}=\sqrt{2}a^{3}M(3+\alpha)^{3/2}\phi-(12H_{0}^{2}\Omega_{M})/a^{3}\\
&V(\phi)=-2M\bigg(\frac{\mathcal{Y}_{1}}{\sqrt{2}M(3+\alpha)}+\frac{\phi\bigg[-\frac{2a^{3}M^{3}\omega^{2}(3+\alpha)^{9/2}}{\mathcal{Y}_{3}((3+\alpha)^{3}\phi+3\mathcal{Y}_{2})^{2}-\frac{\sqrt{2}a^{3}M^{3}(3+\alpha)^{5/2}\big(-12H_{0}^{2}(3+\alpha)^{2}\Omega_{M}+\frac{2a^{3}M\omega^{2}\sqrt{3+\alpha}}{(3+\alpha)^{3}\phi+3\mathcal{Y}_{2}}\big)}{(\mathcal{Y}_{3})^{2}}}\bigg]}{2\sqrt{2}M(3+\alpha)\mathcal{Y}_{1}}\bigg)^{2}\\
&\times\cosh^{2}\big[\frac{\phi \mathcal{Y}_{1}}{\sqrt{2}M(3+\alpha)}\big]+3M^{2}\sinh^{2}\big[\frac{\mathcal{Y}_{1}}{\sqrt{2}M(3+\alpha)}\big],
\end{split}
\end{equation}

and

\begin{equation}\label{30}
\begin{split}
&\mathcal{Y}_{1}=\bigg(\frac{M^{2}(3+\alpha)\bigg[-12H_{0}^{2}(3+\alpha)^{2}\Omega_{M}+\frac{2a^{3}M\sqrt{3+\alpha}\omega^{2}}{(3+\alpha)^{3}\phi+3\sqrt{\frac{\sqrt{2}a^{3}M(3+\alpha)^{2}-3H_{0}^{2}\Omega_{M}}{a^{3}}}}\bigg]}{\sqrt{2}a^{3}M(3+\alpha)^{3/2}\phi-12H_{0}^{2}\Omega_{M}/a^{3}}\bigg)^{\frac{1}{2}}\\
&\mathcal{Y}_{2}=\bigg((\sqrt{2}a^{3}M(3+\alpha)^{2}-3H_{0}^{2}\Omega_{M})/a^{3}\bigg)^{1/2}\\
&\mathcal{Y}_{3}=\sqrt{2}a^{3}M(3+\alpha)^{3/2}\phi-(12H_{0}^{2}\Omega_{M})/a^{3}\\
&\dot{\phi}=-M\bigg(\frac{\mathcal{Y}_{1}}{\sqrt{2}M(3+\alpha)}+\frac{\phi\bigg[-\frac{2a^{3}M^{3}\omega^{2}(3+\alpha)^{9/2}}{\mathcal{Y}_{3}((3+\alpha)^{3}\phi+3\mathcal{Y}_{2})^{2}-\frac{\sqrt{2}a^{3}M^{3}(3+\alpha)^{5/2}\big(-12H_{0}^{2}(3+\alpha)^{2}\Omega_{M}+\frac{2a^{3}M\omega^{2}\sqrt{3+\alpha}}{(3+\alpha)^{3}\phi+3\mathcal{Y}_{2}}\big)}{(\mathcal{Y}_{3})^{2}}}\bigg]}{2\sqrt{2}M(3+\alpha)\mathcal{Y}_{1}}\bigg)\\
&\times\cosh\big[\frac{\phi \mathcal{Y}_{1}}{\sqrt{2}M(3+\alpha)}\big]-\bigg\{-\frac{\omega^{2}}{a^{6}\phi^{2}}-\frac{3H_{0}^{2}\Omega_{M}}{a^{3}}\\
&+M\bigg[\frac{\phi\bigg(\frac{4a^{3}M^{3}(3+\alpha)^{15/2}\omega^{2}}{\mathcal{Y}_{3}\big((3+\alpha)^{3}\phi+3\mathcal{Y}_{2}\big)^{3}}+\frac{4\sqrt{2}a^{6}M^{4}(3+\alpha)^{6}\omega^{2}}{(\mathcal{Y}_{3})^{2}\big((3+\alpha)^{3}\phi+3\mathcal{Y}_{2}\big)^{2}}+\frac{4a^{6}M^{4}(3+\alpha)^{4}\big[-12H_{0}^{2}\Omega_{M}(3+\alpha)^{2}+\frac{2a^{3}M\sqrt{3+\alpha}\omega^{2}}{(3+\alpha)^{3}\phi+3\mathcal{Y}_{2}}\big]}{(\mathcal{Y}_{3})^{3}}\bigg)}{2\sqrt{2}M(3+\alpha)\mathcal{Y}_{1}}\\
&+\frac{-\frac{2a^{3}M^{3}\omega^{2}(3+\alpha)^{9/2}}{\mathcal{Y}_{2}\big((3+\alpha)^{3}\phi+3\mathcal{Y}_{2}\big)^{2}}-\frac{\sqrt{2}a^{3}M^{3}(3+\alpha)^{5/2}\big(-12H_{0}^{2}\Omega_{M}(3+\alpha)^{2}+\frac{2a^{3}M\sqrt{3+\alpha}\omega^{2}}{(3+\alpha)^{3}\phi+3\mathcal{Y}_{2}}\big)}{(\mathcal{Y}_{2})^{2}}}{\sqrt{2}M(3+\alpha)\mathcal{Y}_{1}}\\
&-\frac{\phi\bigg(-\frac{2a^{3}M^{3}\omega^{2}(3+\alpha)^{9/2}}{\mathcal{Y}_{3}\big((3+\alpha)^{3}\phi+3\mathcal{Y}_{2}\big)^{2}}-\frac{\sqrt{2}a^{3}M^{3}(3+\alpha)^{5/2}\big[-12H_{0}^{2}\Omega_{M}(3+\alpha)^{2}+\frac{2a^{3}M\sqrt{3+\alpha}\omega^{2}}{(3+\alpha)^{3}\phi+3\mathcal{Y}_{2}}\big]}{(\mathcal{Y}_{3})^{2}}\bigg)^{2}}{4\sqrt{2}M(\mathcal{Y}_{1})^{3}}\bigg]\cosh\big[\frac{\phi \mathcal{Y}_{1}}{\sqrt{2}M(3+\alpha)}\big]\\
&+M\bigg(\frac{\mathcal{Y}_{1}}{\sqrt{2}M(3+\alpha)}+\frac{\phi\bigg(-\frac{2a^{3}M^{3}\omega^{2}(3+\alpha)^{9/2}}{\mathcal{Y}_{3}\big((3+\alpha)^{3}\phi+3\mathcal{Y}_{2}\big)^{2}}-\frac{\sqrt{2}a^{3}M^{3}(3+\alpha)^{5/2}\big[-12H_{0}^{2}\Omega_{M}(3+\alpha)^{2}+\frac{2a^{3}M\sqrt{3+\alpha}\omega^{2}}{(3+\alpha)^{3}\phi+3\mathcal{Y}_{2}}\big]}{(\mathcal{Y}_{3})^{2}}\bigg)}{2\sqrt{2}M(3+\alpha)\mathcal{Y}_{1}}\bigg)^{2}\\
&\times\sinh\big[\frac{\mathcal{Y}_{1}}{\sqrt{2}M(3+\alpha)}\big]\bigg\},
\end{split}
\end{equation}

\section{Refined swampland conjecture}

This section provides a brief description of the swampland program, especially the refined swampland conjecture.
Then, using the exact solutions obtained for the potential using the anisotropic constant-roll condition concerning complex quintessence, we combine them with swampland conjectures and explain the results by plotting some figures.
We will also examine the compatibility or incompatibility of the mentioned concepts.
Recently, researchers have introduced a concept called the swampland program, which has many conjectures. These conjectures have been studied in various cosmological structures and string theory, and their results have been compared with the latest observable data.
Although this concept is a developing topic and still faces many problems, it has also yielded exciting results.
Also, many conjectures have been added to this concept in recent years, and even conjectures that contradict specific theories have been corrected, including refined swampland conjecture. We also challenge this conjecture in this article, which is expressed in the following form.

\begin{equation}\label{31}
|\nabla V|\geq\frac{C}{M_{p}}V, \hspace{12pt}  min \nabla\partial V\leq -\frac{C '}{M_{pl}^{2}}V
\end{equation}

The above equations for the $V>0$, which is given by

\begin{equation}\label{32}
\sqrt{2\epsilon_{V}}\geq C ,\hspace{12pt}  or \hspace{12pt}\eta_{V}\leq -C '
\end{equation}

\subsection{discussion and result}

We plotted the swampland conjectures for the complex form of the scalar field concerning the exact solutions calculated for the potential with the anisotropic constant-roll condition for two different boundary conditions in figures (1) and (2).
Figure (1) shows the first boundary condition, according to equations (26), (31), and Figure (2) is also related to boundary conditions 2 and is plotted according to equations (29) and (31).
 In figure (1),  we determine the changes of the swampland conjecture are plotted in terms of the complex form of the scalar field and the constant parameters mentioned in the text.
The virtual range considered for the first and second components of the swampland conjecture viz $C$ and $C'$ is 0 to 1.
So the values of second component $C'$ smaller than the $C$.
For this reason, in figures (1b) and (1d), we limited our figures to acceptable values for these swampland components. The allowable range of the two components for the first boundary condition is given according to the constant mentioned parameters and various values of $\alpha$.
 For each of the two swampland components (C, C'), $\alpha=-2$ shows a more acceptable spectrum than the other values.
Of course, by changing the constant parameters, the final results will be almost the same, and there will be some changes in the range of swampland components that can be ignored.
As mentioned in the text, figures are plotted for positive potential.
However, Negative values of potential had almost similar results, which we declined. Of course, the computational process is exactly like this path but with the opposite sign.
Figure (2) is also plotted according to the second boundary condition.
According to all the mentioned points, the allowable range components of the swampland conjectures are specified in figures (2b) and (2d).
As clear in these figures, the acceptable range for the mentioned model is more volume than the previous case.
Similar to the previous case, We plot the figures according to the mentioned constant parameters\cite{40,41} and different parameter $\alpha$ .
 Here, $\alpha=-1,-2$ gives us a more acceptable range and is therefore similar to the previous case.
Of course, the allowable range for each of the two components $(C, C ')$ is beyond the previous case. in general, it can be concluded that compatibility between different concepts as anisotropic constant-roll condition, swampland conjectures, and complex quintessence is observed. The calculations of this section can be expanded and compared with other possible conditions.

\begin{figure}[h!]
 \begin{center}
 \subfigure[]{
 \includegraphics[height=5.5cm,width=5.5cm]{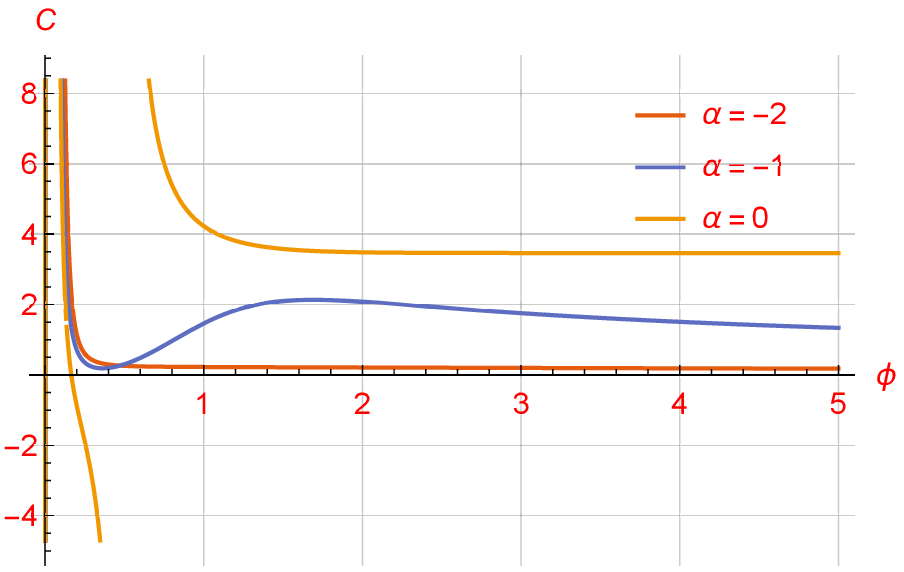}
 \label{1a}}
 \subfigure[]{
 \includegraphics[height=5.5cm,width=5.5cm]{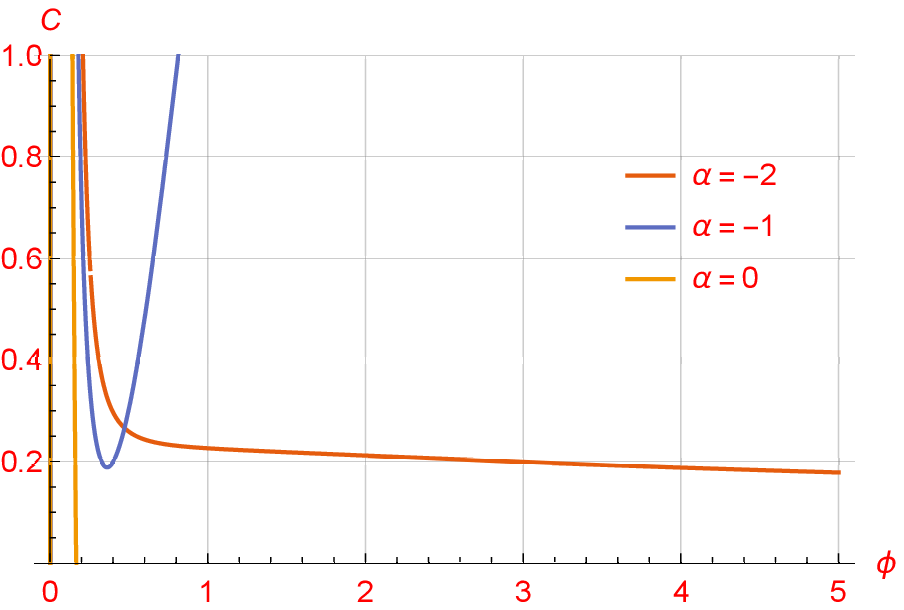}
 \label{1b}}
 \subfigure[]{
 \includegraphics[height=5.5cm,width=5.5cm]{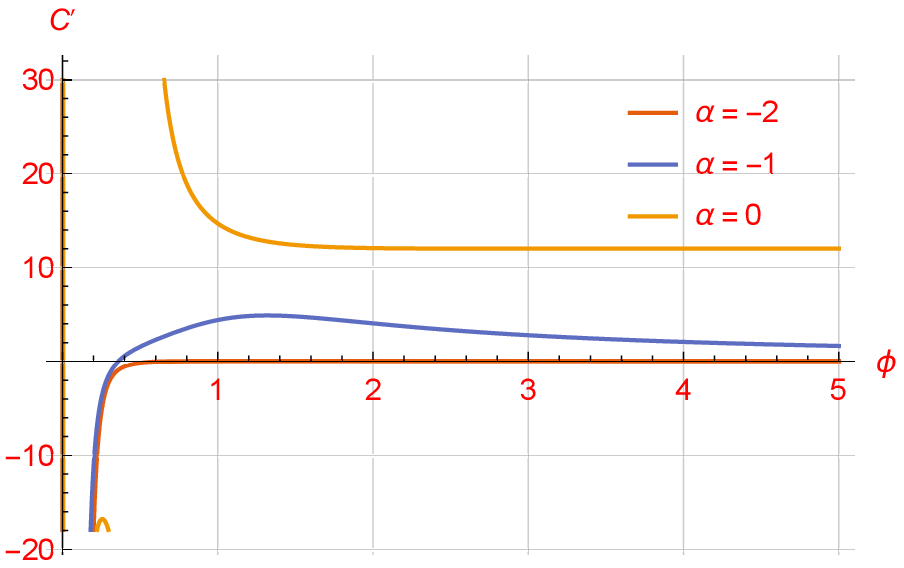}
 \label{1c}}
 \subfigure[]{
 \includegraphics[height=5.5cm,width=5.5cm]{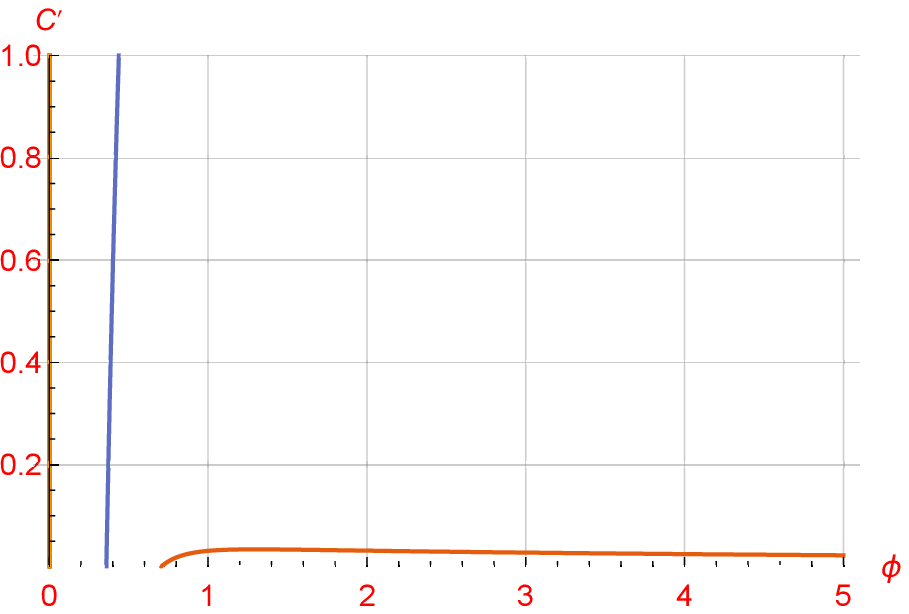}
 \label{1d}}
  \caption{\small{The plot of $C$ and $C'$ in term of $\phi$ for the case I in fig (1a, 1b) and fig (1c, 1d)  respectively, concerning constant mentioned parameters $\omega=1000$, $\Omega_{M}=0.3$, $H_{0}=67.9$, $M=1$ and various values of $\alpha$}}
 \label{1}
 \end{center}
 \end{figure}

\begin{figure}[h!]
 \begin{center}
 \subfigure[]{
 \includegraphics[height=5.5cm,width=5.5cm]{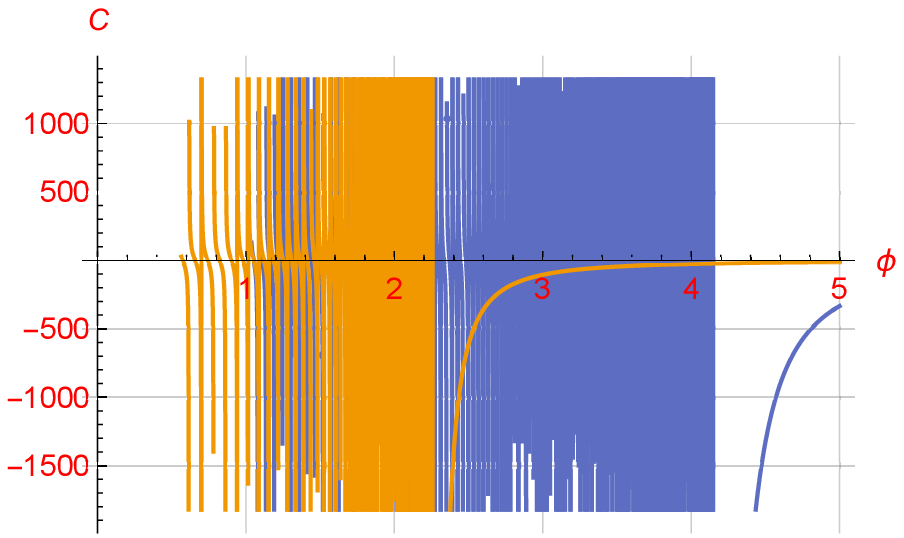}
 \label{2a}}
 \subfigure[]{
 \includegraphics[height=5.5cm,width=5.5cm]{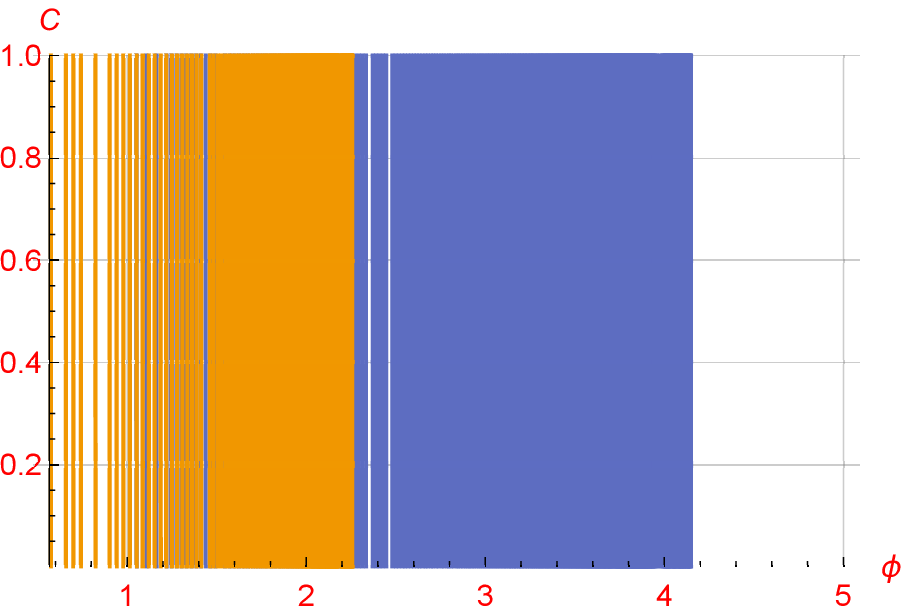}
 \label{2b}}
 \subfigure[]{
 \includegraphics[height=5.5cm,width=5.5cm]{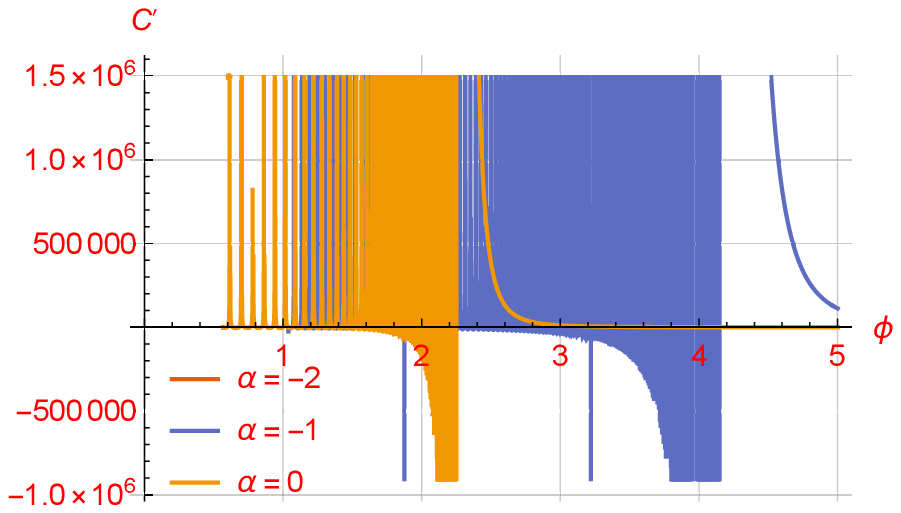}
 \label{2c}}
 \subfigure[]{
 \includegraphics[height=5.5cm,width=5.5cm]{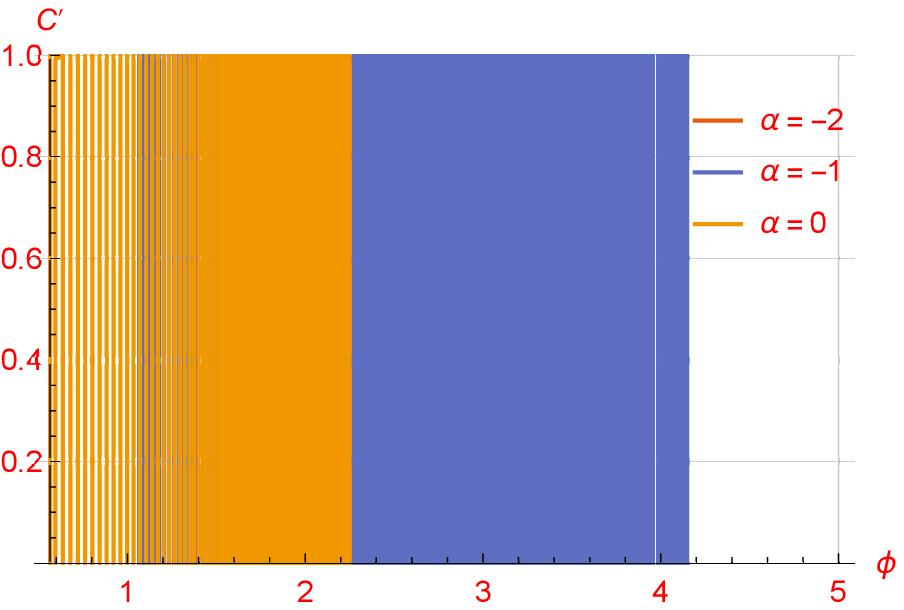}
 \label{2d}}
  \caption{\small{The plot of $C$ and $C'$ in term of $\phi$ for the case II in fig (2a, 2b) and fig (2c, 2d)  respectively, concerning constant mentioned parameters $\omega=1000$, $\Omega_{M}=0.3$, $H_{0}=67.9$, $M=1$ and various values of $\alpha$}}
 \label{1}
 \end{center}
 \end{figure}
\section{Concluding remarks}
Regarding the observations of type Ia supernovae indicating a Universe accelerating expansion, we considered the complex form of the scalar field, called the complex quintessence field, to explain this acceleration with a specific inflationary scenario. We investigated the anisotropic constant-roll of an inflationary scenario using the complex quintessence in the presence of swampland conjectures. The recently introduced complex quintessence has yielded exciting results in the accelerating expansion of the universe connecting with various conditions.
Therefore, using the anisotropic constant-roll conditions, the Lagrangian density of the quintessence field, which leads to obtaining the field equations of the complex scalar field, we calculate the exact solutions for some cosmology parameters such as Hubble parameter, potential, etc.
Using the exact solution of potential and refined swampland conditions, we plot some figures that challenge the three concepts; complex quintessence, anisotropic constant-roll conditions, and swampland conjectures. Then we discuss their compatibility and incompatibility with corresponding results. Finally, we analyzed the complex quintessence in examining the inflationary scenario. We analyzed the results of each boundary condition concerning the exact solutions of the potentials arising from the anisotropic constant-roll condition relating to the complex form of the scalar field as complex quintessence and the composition of the swampland conjectures. We described the results in detail in the previous section.
There are also new questions in this area worth pondering and can be explored in future work.\\
First, is there a complex form for other fields like phantom and Tachyonic? If any, one can calculate the field equations, examine them with an anisotropic constant-roll condition, and compare the results with this paper.\\
Secondly, the results of this research can be examined in the non-commutative space, which can lead to exciting results.\\
It is also possible to study other cosmological concepts of this model and compare the results with the latest observable data by obtaining other cosmological parameters.


\begin{thebibliography}{11}
\bibitem{1}
S. Perlmutter et al, Astrophys. J. 517 565 (1999).
\bibitem{2}
A. G. Riess et al, Astron. J. 116  1009(1998).
\bibitem{3}
P. de Bernardis et al., Nature 404 955 (2000).
\bibitem{4}
S. Hanany et al. Astrophys. J. 545 L5 (2000).
\bibitem{5}
P. de Bernardis et al., astro-ph/0011469.
\bibitem{6}
A. Balbi et al., Astrophys. J. 545 (2000).
\bibitem{7}
P. J. E. Peebles and B. Ratra, Astrophys. J., 325 L17 (1988).;
J. P. Ostriker and P. J. Steinhardt, Nature 377 600 (1995).; A. R. Liddle, D. H. Lyth,
P. T. Viana and M. White, Mon. Not. Roy. Astron. Soc. 282 281 (1996).
C. Armendariz-Picon, V. F. Mukhanov and P. J. Steinhardt, Phys. Rev. D, 63 103510 (2001);
R. R. Caldwell, Phys. Lett. B 545 23 (2002);
J. A. Frieman, C. T. Hill, A. Stebbins and I. Waga, Phys. Rev. Lett., 75 2077 (1995).;
C. Armendariz-Picon, V. F. Mukhanov and P. J. Steinhardt, Phys. Rev. D, 63 103510 (2001);
F. Piazza and S. Tsujikawa, JCAP, 0407 004 (2004).;
R. S. Nojiri and S. D. Odintsov, Phys. Lett. B 562 147 (2003);
B. Ratra and P. J. E. Peebles, Phys. Rev. D, 37 3406 (1988).;
M. S. Turner and M. J. White, Phys. Rev. D, 56 4439 (1997).;
R. R. Caldwell, R. Dave and P. J. Steinhardt, Phys. Rev. Lett., 80 1582 (1998).;
C. Armendariz-Picon, V. F. Mukhanov and P. J. Steinhardt, Phys. Rev. Lett., 85 4438 (2000);
C. Wetterich, Nucl. Phys. B, 302 668 (1988).;
L. M. Krauss and M. S. Turner, Gen. Rel. Grav. 27 1137 (1995).;
M. S. Turner and M. J. White, Phys. Rev. D, 56 4439 (1997).;
R. R. Caldwell, Phys. Lett. B 545 23 (2002);
R. R. Caldwell, M. Kamionkowski and N. N. Weinberg, Phys. Rev. Lett. 91 071301 (2003);
R. S. Nojiri and S. D. Odintsov, Phys. Lett., B 565 1 (2003);
A. Sen, JHEP, 0207 (2002) 065.;
R. R. Caldwell, Phys. Lett. B 545 23 (2002);
N. Arkani-Hamed, H. C. Cheng, M. A. Luty and S. Mukohyama, JHEP 0405 074 (2004).;
F. Piazza and S. Tsujikawa, JCAP, 0407 004 (2004).;
L. M. Krauss and M. S. Turner, Gen. Rel. Grav. 27 1137 (1995).;
M. S. Turner and M. J. White, Phys. Rev. D, 56 4439 (1997).;
R. R. Caldwell, R. Dave and P. J. Steinhardt, Phys. Rev. Lett., 80 1582 (1998).;
C. Armendariz-Picon, V. F. Mukhanov and P. J. Steinhardt, Phys. Rev. Lett., 85 4438 (2000);
L. M. Krauss and M. S. Turner, Gen. Rel. Grav. 27 1137 (1995).;
N. Arkani-Hamed, H. C. Cheng, M. A. Luty and S. Mukohyama, JHEP 0405 074 (2004).;
F. Piazza and S. Tsujikawa, JCAP, 0407 004 (2004).;
\bibitem{8}
R. R. Caldwell, R. Dave and P. J. Steinhardt, Phys. Rev. Lett. 80  1582(1998).;
B. Feng, X. L. Wang and X. M. Zhang, Phys. Lett. B, 607 35 (2005);
Z. K. Guo, Y. S. Piao, X. M. Zhang and Y. Z. Zhang, Phys. Lett. B, 608 177 (2005);
X. Zhang, Commun. Theor. Phys., 44 762 (2005);
A. Anisimov, E. Babichev and A. Vikman, JCAP, 0506 006 (2005);
E. Elizalde , S. Nojiri, and S. D. Odintsov, Phys. Rev. D, 70 043539 (2004);
S. Nojiri, S. D. Odintsov, and S. Tsujikawa, Phys. Rev. D, 71 063004 (2005);
C. Deffayet, G. R. Dvali and G. Gabadadze, Phys. Rev. D 65 044023 (2002).;
L. Amendola, Phys. Rev. D, 62 043511 (2000).;
A. Y. Kamenshchik, U. Moschella and V. Pasquier, Phys. Lett. B 511 265 (2001).
\bibitem{9}
I. Zlatev, L. Wang and P. J. Steinhardt, Phys. Rev. Lett. 82 896(1999).; P. J. Steinhardt, L. Wang and I. Zlatev, Phys. Rev. D 59 123504 (1999).
\bibitem{10}
D. Huterer and M. S. Turner, Phys. Rev. D 60 081301 (1999).
\bibitem{11}
A. A. Starobinsky, JETP Lett. 68 (1998) 757 [Pisma Zh. Eksp. Teor. Fiz. 68 721 (1998).
\bibitem{12}
T. Chiba and T. Nakamura, Phys. Rev. D 62 121301 (2000).
\bibitem{13}
J. Sadeghi, and S. Noori Gashti, \textit{Tsallis holographic dark energy under Complex form of Quintessence model} (2022).
\bibitem{14}
J. Sadeghi, and S. Noori Gashti, \textit{Tsallis and Kaniadakis holographic dark energy with Complex Quintessence theory in Brans–Dicke cosmology} (2022).
\bibitem{15}
Yang Liu, Eur. Phys. J. C 80 1204 (2020).
\bibitem{16}
H. Motohashi, A.A. Starobinsky, J. Yokoyama, JCAP 1509 (2015).
\bibitem{17}
H. Motohashi, A.A. Starobinsky, JCAP 11, 025 (2019).
\bibitem{18}
H. Motohashi, A.A. Starobinsky, EPL 117 (2017).
\bibitem{19}
L. Anguelova, P. Suranyi, L.C.R. Wijewardhana, JCAP 02, 004 (2018).
\bibitem{20}
Mehdi Shokri, Jafar Sadeghi, Saeed Noori Gashti, Physics of the Dark Universe, 100923 (2021).
\bibitem{21}
T.Q. Do, W.F. Kao, I.C. Lin, Phys. Rev. D 83, 123002 (2011).;
K. Murata, J. Soda, JCAP 1106, 037 (2011).;
M.A. Watanabe, S. Kanno, J. Soda, Phys. Rev. Lett. 102, 191302 (2009).;
J. Soda, Class. Quantum Gravity 29, 083001 (2012)
A. Maleknejad, M.M. Sheikh-Jabbari, J. Soda, Phys. Rep. 528,
161 (2013).;
A. Ito, J. Soda, Phys. Rev. D 92(12), 123533 (2015).;
R. Emami, arXiv:1511.01683 (2015).;
J. Sadeghi, S. Noori Gashti, Eur. Phys. J. C 81, 301 (2021).;
S. Kanno, J. Soda, M.A. Watanabe, JCAP 12, 024 (2010).;
J. Ohashi, J. Soda, S. Tsujikawa, JCAP 1312, 009 (2013).;
K. Yamamoto, M.A. Watanabe, J. Soda, Class. Quantum Gravity
29, 145008 (2012).;
J. Ohashi, J. Soda, S. Tsujikawa, Phys. Rev. D 87(8), 083520
(2013).;
K. Yamamoto, Phys. Rev. D 85, 123504 (2012).;
T.Q. Do, W.F. Kao, Phys. Rev. D 84, 123009 (2011).;
M. Thorsrud, D.F. Mota, S. Hervik, JHEP 1210, 066 (2012).;
J. Ohashi, J. Soda, S. Tsujikawa, Phys. Rev. D 88, 103517 (2013).;
S. Hervik, D.F. Mota, M. Thorsrud, JHEP 1111, 146 (2011).;
R. Emami, H. Firouzjahi, S.M. SadeghMovahed, M. Zarei, JCAP
1102, 005 (2011).;
A. Ito, J. Soda, JCAP 1604(04), 035 (2016).;
M. Karciauskas, Mod. Phys. Lett. A 31(21), 1640002 (2016).;
S. Lahiri, JCAP 1609(09), 025 (2016).;
N. Bartolo, S. Matarrese, M. Peloso, A. Ricciardone, JCAP 1308,
022 (2013).;
K.I. Maeda, K. Yamamoto, Phys. Rev. D 87(2), 023528 (2013).;
K.I. Maeda, K. Yamamoto, JCAP 1312, 018 (2013).;
B. Chen, Z.W. Jin, JCAP 1409(09), 046 (2014).;
S. Lahiri, JCAP 1701(01), 022 (2017).;
M. Adak, O. Akarsu, T. Dereli, O. Sert, JCAP 11, 026 (2017).;
M. Tirandari, K. Saaidi, Nucl. Phys. B 925, 403 (2017).;
T.Q. Do, W.F. Kao, Phys. Rev. D 96(2), 023529 (2017).;
A.E. Gumrukcuoglu, B. Himmetoglu, M. Peloso, Phys. Rev. D
81, 063528 (2010).;
T.Q. Do, S.H.Q. Nguyen, Int. J. Mod. Phys. D 26(07), 1750072
(2017).;
T.R. Dulaney, M.I. Gresham, Phys. Rev. D 81, 103532 (2010).;
\bibitem{22}
C. Vafa, arXiv:hep-th/0509212 (2005).;
Kenji Kadota, Chang Sub Shin, Takahiro Terada and Gansukh Tumurtushaa, JCAP01 008 (2020).
;Oikonomou, V. K. arXiv:2012.01312 (2020).
\bibitem{23}
H. Ooguri, C. Vafa, Nucl. Phys. B 766, 21–33 (2007).;
Oem Trivedi, arXiv:2101.00638 (2021).; Suratna Das, physics of the dark univers 27, 100432 (2020).
\bibitem{24}
N. Arkani-Hamed, L. Motl, A. Nicolis, C. Vafa, JHEP 06, 060 (2007).;
Abolhassan Mohammadi, Tayeb Golanbari, Jamil Enayati, arXiv:2012.01512, (2020).
\bibitem{25}
J. Sadeghi, S.N. Gashti Pramana 95 (198) (2021).; J. Sadeghi, B. Pourhassan, S.N. Gashti, S. Upadhyay
Physica Scripta 96 (12), 125317 (2021).; J. Sadeghi, B. Pourhassan, S.N. Gashti, S. Upadhyay
arXiv preprint arXiv:2201.04071 (2022).
\bibitem{26}
M. Shokri, J. Sadeghi, R. Herrera, S.N. Gashti arXiv preprint arXiv:2112.12309 (2021).
\bibitem{27}
S.N. Gashti, Journal of Holography Applications in Physics 2 (1), 13-24 (2022)
\bibitem{28}
M. Orellana, F. Garcia, F. Teppa Pannia, G. Romero, Gen. Relativ. Gravit. 45, 771–783 (2013).;
Osses, Constanza and Videla, Nelson and Panotopoulos, Grigoris, arXiv:2101.08882, (2021).;Suddhasattwa Brahma, Phys.Rev.D 101  2, 023526 (2020).; Robert Brandenberger,  arXiv:2102.09641 (2021).
\bibitem{29}
S. Nojiri, S.D. Odintsov, Phys. Rep. 505, 59 (2011).;
William H. Kinney, Phys. Rev. Lett.122, 8 081302 (2019).;William H. Kinney, arXiv:2103.16583 (2021).; Yu, Ten-Yeh and Wen, Wen-Yu, Phys. Lett. B781, 713--718  (2018).; Yang Liu, Eur. Phys. J. C 81, 1122 (2021).
\bibitem{30}
S. Capozziello, M. De Laurentis, S.D. Odintsov, A. Stabile, Phys. Rev. D 83, 064004 (2011).
\bibitem{31}
S. Capozziello, M. Faizal, M. Hameeda, B. Pourhassan, V. Salzano, S. Upadhyay, Mon. Not. R. Astron. Soc. 474, 2430–2443 (2018).
\bibitem{32}
A. Arapoglu, C. Deliduman, K.Y. Eksi, JCAP 1107, 020 (2011).
\bibitem{33}
S. Capozziello, R. D’Agostino, O. Luongo, Int. J. Mod. Phys. D 28, 1930016 (2019).
\bibitem{34}
S. Capozziello, R. D’Agostino, O. Luongo, JCAP 1805, 008 (2018).
\bibitem{35}
S. Capozziello, R. D’Agostino, O. Luongo, Gen. Relativ. Gravit. 51, 2 (2019).
\bibitem{36}
P. Channuie, Eur. Phys. J. C 79, 508 (2019).
\bibitem{37}
J. Sadeghi, E.N. Mezerji, S.N. Gashti, Modern Physics Letters A 36 (05), 2150027 (2021).
\bibitem{38}
R. Myrzakulov, L. Sebastian, S. Vagnozzi, Eur. Phys. J. C 75, 444 (2015).
\bibitem{39}
J. Gu, and W. Hwang, Phys. Lett. B, 517 1 (2001).
\bibitem{40}
S. Ghaffari, arXiv:2112.05813 (2021).
\bibitem{41}
S. Ghaffari, et al., J. C  78, 706 (2018).
\end{thebibliography}
\end{document}